\documentclass[twocolumn]{IEEEtran}
\usepackage[utf8]{inputenc}
\usepackage{amsmath,amssymb,amsthm}
\usepackage{color,graphics}

\newtheorem{lemma}{Lemma}
\newtheorem{theorem}[lemma]{Theorem}
\newtheorem{corollary}[lemma]{Corollary}
\theoremstyle{definition}
\newtheorem{remark}[lemma]{Remark}

\renewcommand\leq{\leqslant}
\renewcommand\geq{\geqslant}

\title{New Lower Bounds on the Generalized Hamming Weights of AG Codes}

\author{Maria Bras-Amor\'os, Kwankyu Lee, and Albert Vico-Oton
\thanks{M.~Bras-Amor\'os is with the Department of Computer Engineering and Mathematics, Universitat Rovira i Virgili, Tarragona 43007, Catalonia, Spain (e-mail: maria.bras@urv.cat). She was partly supported by the Spanish Government through project TIN2012-32757 ``ICWT'', and CONSOLIDER INGENIO 2010 CSD2007-0004 ``ARES'',  and by the Government of Catalonia under grant 2014 SGR 537.}
\thanks{K.~Lee is with the Department of Mathematics and Education, Chosun University, Gwangju 501-759, Korea (e-mail: kwankyu@chosun.ac.kr). He was supported by Basic Science Research Program through the National Research Foundation of Korea (NRF) funded by the Ministry of Education, Science and Technology (2013R1A1A2009714).}%
\thanks{Albert Vico-Oton is with the Department of Computer Engineering and Mathematics, Universitat Rovira i Virgili, Tarragona 43007, Catalonia, Spain (e-mail: albert.vico@urv.cat). He was partly supported by the Spanish Government through project TIN2012-32757 ``ICWT'', and CONSOLIDER INGENIO 2010 CSD2007-0004 ``ARES'',  and by the Government of Catalonia under grant 2009 SGR 1135.}%
\thanks{A small part of the results presented in this paper
(Section II and a reduced version of Theorem 11) was presented in the
Forum "Galois Geometries and Applications" held at the Royal Flemish
Academy of Belgium for Science and the Arts in 2012 \cite{galois}.}
}

\begin{document}
\maketitle

\begin{abstract}
A sharp upper bound for the maximum integer not belonging to an ideal of a numerical semigroup is given and the ideals attaining this bound are characterized. Then the result is used, through the so-called Feng-Rao numbers, to bound the generalized Hamming weights of algebraic-geometry codes. This is further developed for Hermitian codes and the codes on one of the Garcia-Stichtenoth towers, as well as for some more general families.
\end{abstract}


\section{Introduction}
The generalized Hamming weights of a linear code are the minimum size of the support of the linear subspaces of the code of each given dimension. They have many applications in a variety of fields of communications. The notion was first used by Wei \cite{Wei} to analyze the performance of the wire-tap channel of type II introduced in \cite{OzarowWyner} and in connection to $t$-resilient functions. See also \cite{Munuera}. The connections with the wire-tap channel have been updated recently in \cite{RoSoSp}, this time using network coding. The notion itself has also been generalized for network coding in \cite{NgaiYeungZhang}. The generalized Hamming weights have also been used in the context of list decoding \cite{Guruswami2003,GoGuRa}. In particular, Guruswami shows that his $(e,L)$-list decodibility concept for erasures is equivalent with the generalized Hamming weights for linear codes.
Finally, the generalized Hamming weights also appear for bounding the covering radius of linear codes \cite{JanwaLal}, and recently for secure secret sharing based on linear codes \cite{sssssitw,sssssallerton}.

In this contribution, we deal with the generalized Hamming weights of one-point AG codes from the perspective of the associated Weierstrass semigroup, that is, the set of pole orders of the rational functions having a unique pole at the defining one point. A numerical semigroup is a subset of the nonnegative integers ${\mathbb N}_0$ that contains $0$, is closed under addition, and has a finite complement in ${\mathbb N}_0$. The elements in this complement are called the gaps of the semigroup and the number of gaps is called the genus. The maximum gap is usually referred to as the Frobenius number of the semigroup and the conductor is the Frobenius number plus one. By the pigeonhole principle it is easy to prove that the Frobenius number is at most twice the genus minus one, and there are semigroups, called symmetric semigroups, attaining this bound.

An ideal of a numerical semigroup is a subset of the semigroup such that any element in the subset plus any element of the semigroup add up to an element of the subset. Again the ideal will be a subset of ${\mathbb N}_0$ with finite complement in it. Our first result is an analogue of the upper bound on the Frobenius number of the semigroup, for the largest integer not belonging to an ideal, which will also be called the Frobenius number of the ideal. Indeed, we prove that it is at most the size of the complement of the ideal in the semigroup plus twice the genus minus one (Theorem~\ref{t:maxidealgap}). This generalizes the bound on the Frobenius number of the semigroup since that bound can be derived from this bound by taking the ideal to be the whole semigroup. Then we characterize the ideals whose Frobenius number attains the bound. It turns out that the set of codes in a sequence of one-point AG codes are pairwise isometric to the set of duals of the same codes if and only if the set of pole-orders defining the codes is exactly the complement of one such ideal \cite{GMRT}.

A nice tool for tackling the generalized Hamming weights for AG codes are the generalized order bounds introduced in \cite{HePe}, involving Weierstrass semigroups. In \cite{FaMu}, a constant depending only on the semigroup and the dimension of the Hamming weights was introduced, from which the order bounds could be completely determined for codes of rate low enough. This constant was called Feng-Rao number.
In the present contribution, using the upper bound on the Frobenius number of an ideal, we derive a lower bound on the Feng-Rao numbers and consequently a new bound on the generalized Hamming weights (Theorem~\ref{t:boundEr}, Corollary~\ref{c:bound}). This is done by analyzing the intervals of consecutive gaps of the Weierstrass semigroup. Consecutive gaps were already used in \cite{GaKiLa} for bounding the minimum distance of codes and in \cite{Lizhong} for bounding the generalized Hamming weights, in this case for primal codes. In the last section, we study the intervals of consecutive gaps for Hermitian codes and for codes in one of the Garcia-Stichtenoth towers of codes attaining the 
Drinfeld-Vl\u adu\c t bound, as well as for their respective generalizations to semigroups generated by intervals and inductive semigroups.

\section{The Frobenius number of an ideal}

From now on, $\Lambda$ will denote a numerical semigroup and the elements of $\Lambda$ are denoted $\{\lambda_0=0<\lambda_1<\dots\}$. The Frobenius number is $F$, the conductor is $c$, and the genus is~$g$. Given an ideal $I$ of a numerical semigroup $\Lambda$, we call the size of $\Lambda\setminus I$ the {\it difference} of $I$ with respect to $\Lambda$. We call the ideals of the form $a+\Lambda$ for some $a\in\Lambda$ {\em principal} ideals.
It was proved in \cite[Lemma 5.15]{HoLiPe} that the difference of the principal ideal $a+\Lambda$ is exactly $a$. So, for principal ideals, the Frobenius number of the ideal is at most the difference plus twice the genus of the semigroup minus one. In Theorem~\ref{t:maxidealgap}, we will prove that the same holds for {\it any} ideal of a numerical semigroup. Then we will characterize the semigroups for which the inequality is indeed an equality.

\subsection{An upper bound for the Frobenius number of an ideal}

Define the set of divisors of $\lambda_i$ by $$D(i)=\{\lambda_j\leq \lambda_i:\lambda_i-\lambda_j\in \Lambda\}$$ and $\nu_i=\#D(i)$ for $i\in{\mathbb N}_0$. Some results related to the sequence $\nu_i$ and also to its applications to coding theory can be found for instance in \cite{KiPe,Bras:acute,Bras:anote,MuTo,OnTa2008,OnTa2009,OnTa2010}. Barucci \cite{Barucci} proved the next result.

\begin{lemma}\label{t:Barucci}
Any ideal of a numerical semigroup is an
intersection of irreducible ideals and irreducible ideals
have the form $\Lambda\setminus D(i)$ for some $i$.
\end{lemma}

The next result was proved in \cite[Theorem 5.24]{HoLiPe}.

\begin{lemma}\label{l:KiPe}
Let $g(i)$ be the number of gaps smaller than $\lambda_i$ and $G(i)$ the
number of pairs of gaps adding up to $\lambda_i$. Then
$$\nu_i=i-g(i)+G(i)+1.$$
\end{lemma}

Now we can state the main result of this section.

\begin{theorem}\label{t:maxidealgap}
Suppose that a numerical semigroup has genus $g$. Suppose that $I$ is an ideal of the semigroup with difference $d$. Then, the Frobenius number of $I$ is at most $d+2g-1$. That is, $d+2g+i\in I$ for all $i\geq 0$.
\end{theorem}

\begin{IEEEproof}
If two ideals satisfy the result, then their intersection also satisfies it. So by Lemma~\ref{t:Barucci}, it suffices to prove the result for irreducible ideals. Now we want to prove the result for the ideal $I=\Lambda\setminus D(i)$. That is, $\nu_i+2g\geq \max\{c, \lambda_i+1\}$, where $c$ is the conductor of $\Lambda$. If $c\geq\lambda_i+1$ then we are done since $c\leq 2g$. Suppose then that $\lambda_i+1>c$. Then $g(i)=g$, $\lambda_i=i+g$, and hence by Lemma~\ref{l:KiPe}, $\nu_i+2g=(i-g+G(i)+1)+2g=i+g+1+G(i)=\lambda_i+1+G(i)\geq\lambda_i+1$.
\end{IEEEproof}

\subsection{Ideals attaining the upper bound}

We will devote this section to characterize the ideals of semigroups 
that attain the upper bound on the Frobenius number of the ideal.
We first need some preliminary lemmas.

\begin{lemma}\label{l:Gzcond}
If $G(i)=0$ then $\lambda_i\geq c$.
\end{lemma}

\begin{IEEEproof}
If $G(i)=0$ then, since $1,\dots,\lambda_1-1$ are gaps, 
$\lambda_i-\lambda_1+1,\dots,\lambda_i-1$ are non-gaps.
But also $\lambda_i\in\Lambda$ so the interval 
$[\lambda_i-\lambda_1+1,\dots,\lambda_i]$ is included in $\Lambda$.
Now, by adding multiples of $\lambda_1$ to the elements in this interval we
get the whole set of integers $\lambda_i+k$ with $k\geq 0$. Then $\lambda_i\geq c$.
\end{IEEEproof}

\begin{lemma}\label{l:Gzequiv}
$G(i)=0$ if and only if $\{\lambda_i-F\}\cup\{\lambda_i-F+h: h\not\in\Lambda, F-h\not\in\Lambda\}\subseteq\Lambda.$
\end{lemma}

\begin{IEEEproof}
Suppose $G(i)=0$. Then obviously $\lambda_i-F\in\Lambda$. Now suppose that $h\not\in\Lambda, F-h\not\in\Lambda.$
We need to see that $\lambda_i-F+h\in\Lambda$. But $\lambda_i-F+h=\lambda_i-(F-h)\in\Lambda$ since $G(i)=0$ and $F-h\not\in\Lambda$. On the other hand, suppose that $\{\lambda_i-F\}\cup\{\lambda_i-F+h: h\not\in\Lambda, F-h\not\in\Lambda\}\subseteq\Lambda$ and we want to prove that $G(i)=0$.
If $G(i)\neq 0$ then there exists a gap $h'$ such that $\lambda_i-h'$ is a gap.
But $\lambda_i-h'=(\lambda_i-F)+(F-h')$.
Since $\lambda_i-F\in\Lambda$ by hypothesis, $F-h'$ must be a gap. Let us call this gap $h=F-h'$.
Then both $h$ and $F-h=h'$ are gaps and, 
by the hypothesis, $\lambda_i-F+h\in\Lambda$. But $\lambda_i-F+h=\lambda_i-h'$ is a gap, a contradiction. Then $G(i)=0$.
\end{IEEEproof}

\begin{lemma}\label{l:lsis}
If $G(i)=0$ then $\Lambda\setminus D(i)=\{\lambda_i-h:h\in{\mathbb Z}\setminus \Lambda\}$.
\end{lemma}

\begin{IEEEproof}
By Lemma~\ref{l:Gzcond}, we know that $\lambda_i\geq c$. To see the inclusion $\supseteq$ suppose that $h\in{\mathbb Z}\setminus \Lambda$. If $h<0$ then $\lambda_i-h>\lambda_i$ and thus $\lambda_i\in\Lambda\setminus D(i)$. If $h>0$ then $h<c$ and, since $\lambda_i\geq c$, $\lambda_i-h\geq 0$. Then
$\lambda_i-h\in\Lambda$ because $G(i)=0$. Finally $\lambda_i-h\not\in D(i)$ by definition of $D(i)$. For the reverse inclusion, suppose that $\lambda\in\Lambda\setminus D(i)$.
If $\lambda>\lambda_i$ then $\lambda=\lambda_i-h$ with $h<0$ and so $h\in{\mathbb Z}\setminus\Lambda$.
If $\lambda<\lambda_i$ then $\lambda_i-\lambda$ is a gap $h$ because otherwise $\lambda\in D(i)$. So, 
$\lambda\in\{\lambda_i-h:h\in{\mathbb Z}\setminus\Lambda\}$.
\end{IEEEproof}

\begin{theorem}\label{t:attain}
Suppose that $\Lambda$ is a numerical semigroup of genus $g$. Let $I$ be an ideal of $\Lambda$ with difference $d>0$. Then the next statements are equivalent:
\begin{enumerate}
\item The Frobenius number of $I$ is exactly $d+2g-1$.
\item $I=\Lambda\setminus D(i)$ for some $i$ with $G(i)=0$.
\item $\Lambda\setminus I=\Lambda\cap((d+2g-1)-\Lambda)=\{\lambda\in \Lambda:d+2g-1-\lambda\in\Lambda\}$
\item $I=\{\lambda_i-h: h\in{\mathbb Z}\setminus\Lambda\}$ for some $i$ with $G(i)=0$.
\item  $\{a+h: h\not\in\Lambda, F-h\not\in\Lambda\}\subseteq\Lambda$ 
  and $I=(a+\Lambda)\cup\{a+h: h\not\in\Lambda, F-h\not\in\Lambda\}$ for some $a\in\Lambda$, $a>0$.
\end{enumerate}
\end{theorem}

\begin{IEEEproof}
(1)$\Longleftrightarrow$(2):
Suppose first that $I=\Lambda\setminus D(i)$ for some $i$ with $G(i)=0$. Then $d=\nu_i$. Also, by Lemma~\ref{l:Gzcond}, $g(i)=g$ and $\lambda_i=i+g$. Now, by Lemma~\ref{l:KiPe}, $d+2g-1=\lambda_i\not\in I$.

Conversely, suppose that the Frobenius number of $I$ is $d+2g-1$.
If $I$ is a proper intersection of two ideals $I'$ and $I''$ with difference $d'$ and $d''$ respectively, then $I$ has difference $d$ strictly larger than $d'$ and strictly larger than $d''$. If $d+2g-1$ does not belong to $I$ then it does not belong either to $I'$ or to $I''$, but $d+2g-1$ is strictly larger than $d'+2g-1$  and strictly larger than $d''+2g-1$, contradicting Theorem~\ref{t:maxidealgap}. So, $I$ must be, by Lemma~\ref{t:Barucci}, $\Lambda\setminus D(i)$ for some~$i$.

Since $I=\Lambda\setminus D(i)$, it holds $d=\nu_i$.
If $\lambda_i<c$, then $\nu_i+2g-1\geq 1+2g-1=2g\geq c$ and so $d+2g-1\in I$, which contradicts our assumption.
Therefore $\lambda_i\geq c$. Then
$\nu_i=i-g+G(i)+1$ by Lemma~\ref{l:KiPe}. So $d+2g-1=i+g+G(i)=\lambda_i+G(i)$. 
Since $d+2g-1\not\in I$, it follows that $G(i)=0$.

(2)$\Longleftrightarrow$(3) is immediate by replacing $i$ by $d+g-1$.

(2)$\Longleftrightarrow$(4) follows immediatelly from Lemma~\ref{l:lsis}.

(4)$\Longleftrightarrow$(5) follows from Lemma~\ref{l:Gzequiv},
by setting $a=\lambda_i-F$, and using
the equality $\{\lambda_i-h:h\in{\mathbb Z}\setminus\Lambda\}=\{a+(F-h):h\in{\mathbb Z}\setminus\Lambda\}$, and the fact that
$\{F-h:h\in{\mathbb Z}\setminus\Lambda\}=\Lambda\cup\{h:h\not\in\Lambda,F-h\not\in\Lambda\}$.
\end{IEEEproof}

As an example, consider the semigroup $$\Lambda=\{0,4,5,8,9,10,12,13,\rightarrow\}.$$
We will list all the ideals $I$ satisfying $d+2g-1\not\in I$ ($d$ the difference of $I$).
Since the largest $i$ for which $G(i)>0$ is $16$ as $11+11=22=\lambda_{16}$,
all ideals $I=\Lambda\setminus D(i)$ with $i\geq 17$ attain the bound.
It remains to see what indices $i$ between $6$ and $15$ satisfy $G(i)=0$.

For $i=6$, $G(i)>0$ since $\lambda_i=12=11+1$.

For $i=7$, $G(i)>0$ since $\lambda_i=13=11+2$. 

For $i=8$, $G(i)>0$ since $\lambda_i=14=11+3$.

For $i=9$, $G(i)=0$. Indeed, $\{15-1=14, 15-2=13, 15-3=12, 15-6=9, 15-7=8, 15-11=4\}\subseteq \Lambda$.

For $i=10$ $G(i)=0$. Indeed, $\{16-1=15, 16-2=14, 16-3=13, 16-6=10, 16-7=9, 16-11=5\}\subseteq \Lambda$.
 
For $i=11$ $G(i)>0$ since $\lambda_i=17=11+6$.

For $i=12$ $G(i)>0$ since $\lambda_i=18=11+7$.

For $i=13$ $G(i)=0$. Indeed, $\{19-1=18, 19-2=17, 19-3=16, 19-6=13, 19-7=12, 19-11=8\}\subseteq \Lambda$.

For $i=14$ $G(i)=0$. Indeed, $\{20-1=19, 20-2=18, 20-3=17, 20-6=14, 20-7=13, 20-11=9\}\subseteq \Lambda$.

For $i=15$ $G(i)=0$. Indeed, $\{21-1=20, 21-2=19, 21-3=18, 21-6=15, 21-7=14, 21-11=10\}\subseteq \Lambda$.

Hence, all ideals attaining the bound in Theorem~\ref{t:maxidealgap} are
\[
	I_9=\Lambda\setminus D(9)=\{4,8,9,12,13,14,16,17,18,19,20,21,22,\dots\},
\]
with $D(9)=\{0,5,10,15\},d=4, d+2g-1=15$;
\[
	I_{10}=\Lambda\setminus D(10)=\{5,9,10,13,14,15,17,18,19,20,21,22,\dots\},
\]
with $D(10)=\{0,4,8,12,16\},d=5, d+2g-1=16$;
\[
	I_{13}=\Lambda\setminus D(13)=\{8,12,13,16,17,18,20,21,22,\dots\},
\]
with $D(13)=\{0,4,5,9,10,14,15,19\},d=8, d+2g-1=19$;
\[
	I_{14}=\Lambda\setminus D(14)=\{9,13,14,17,18,19,21,22,\dots\},
\]
with $D(14)=\{0,4,5,8,10,12,15,16,20\}$, $d=9$, $d+2g-1=20$;
\[
	I_{15}=\Lambda\setminus D(15)=\{10,14,15,18,19,20,22,\dots\},
\]
with $D(15)=\{0,4,5,8,9,12,13,16,17,21\}$, $d=10$, $d+2g-1=21$;
\[
	I_{17}=\Lambda\setminus D(17)=\{12,16,17,20,21,22,24,\dots\},
\]
with $D(17)=\{0,4,5,8,9,10,13,14,15,18,19,23\}$, $d=12$, $d+2g-1=23$; and $\Lambda\setminus D(i)$ for all $i> 17$.
In this last case, $D(i)=\{0,4,5,8,9,10,12,13,\dots,i+6-12,i+6-10,i+6-9,i+6-8,i+6-5,i+6-4,i+6\}$, $d=i-5$, $d+2g-1=i+6$.

In the next corollary we prove that for a symmetric semigroup, the ideals attaining the bound on the Frobenius number of the ideal are exactly the principal ideals.
\begin{corollary}
Let $\Lambda$ be a {\it symmetric} numerical semigroup with Frobenius number $F$ and genus $g$. Suppose that $I$ is an ideal of $\Lambda$ with difference $d$. Then the Frobenius number of $I$ is $d+2g-1$ if and only if 
$I$ is principal.
\end{corollary}

\begin{IEEEproof}
It follows from Theorem~\ref{t:attain} and the fact that for any gap $h$ of a symmetric semigroup, $F-h\in\Lambda$.
\end{IEEEproof}

This can be checked again with the previous example since the semigroup $\Lambda$ in there is symmetric.
Notice though that the hypothesis of being symmetric is necessary. For instance, take $\Lambda=\{0,4,8,9,\dots\}$ which has genus $6$ and Frobenius number $7$ and so it is not symmetric. Consider its ideal 
$$I=\Lambda\setminus D(10)=\Lambda\setminus\{0,4,8,12,16\}=\{9,10,11,13,14,15,17,\dots\}$$
Its difference is $d=5$ and its Frobenius number is $d+2g-1=16$. However, $I$ is not $$9+\Lambda=\{9,13,17,18,\dots\}.$$ The elements $10,11,14,15$ have to be included in $I$ in order to have $d+2g-1\not\in I$.
Hence,  $I$ is not principal as $I=(9+\Lambda)\cup\{10,11,14,15\}$.

\begin{remark}
It is shown in \cite{GMRT} that the ideals attaining the bound in Theorem~\ref{t:maxidealgap} arise in 
the characterization of sequences of one-point AG codes that are auto-dual in the following sense.
Two codes $C,D\subseteq{\mathbb F}_q^n$ are said to be $x$-isometric, for $x\in{\mathbb F}_q^n$ if and only if the map $\chi_x:{\mathbb F}_q^n\rightarrow{\mathbb F}_q^n$ given by the component-wise product $\chi_x(v)=x * v$ satisfies
$\chi_x(C)=D$. Then, a sequence of codes $(C_i)_{i={0,\dots,n}}$ is said to satisfy the {\it isometry-dual condition} if there exists $x\in({\mathbb F}_q^*)^n$ such that $C_i$ is $x$-isometric to $C_{n-i}^\perp$ for all $i=0,1,\dots,n$.
Now let $P_1,\dots,P_n, Q$ be different rational points of a (projective, non-singular, geometrically irreducible) curve with genus $g$
and define $C_m=\{(f(P_1),\dots,f(P_n)):f\in L(mQ)\}$.
Note that it can be the case that $C_m=C_{m-1}$.
Let $W$ be the Weierstrass semigroup at $Q$ and let $W^*=\{0\}\cup\{m\in{\mathbb N}, m>0:C_m\neq C_{m-1}\}=\{m_0=0,m_1,\dots,m_n\}$. Then 
$W\setminus W^*$ is an ideal of $W$ (this is stated in different words in \cite[Corollary 3.3.]{GMRT}).
In particular, $C_{m_0},C_{m_1},\dots,C_{m_n}$ satisfies the isometry-dual condition if and only if $n+2g-1\in W^*$, that is, if and only if $W\setminus W^*$ hits the bound in Theorem~\ref{t:maxidealgap}. This is proved in \cite[Proposition 4.3.]{GMRT}.
\end{remark}

\section{A lower bound on the Feng-Rao numbers}

\subsection{Feng-Rao numbers}

Suppose $\Lambda=\{\lambda_0=0<\lambda_1<\dots\}$ is a numerical semigroup.
In coding theory, the $\nu$ sequence of $\Lambda$ defined above is very important. In particular, for an algebraic curve with Weierstrass semigroup $\Lambda$ at a rational point $P$, the order (or Feng-Rao) bound on the minimum distance of the duals of the one-point codes defined on $P$ by the evaluation of rational functions having only poles at $P$ of order at most $\lambda_m$ is defined as $\delta(m)=\min\{\nu_i:i>m\}$ \cite{FeRa:d,KiPe,HoLiPe}. Some results on its computation can be found in \cite{CaFa,HoLiPe,Bras:acute,MuTo,OnTa2008,OnTa2009,OnTa2010}.

A generalization of this bound is the $r$-th order bound on the generalized $r$-th generalized Hamming weight. For this define $D(i)$ as before and 
\[
	D(i_1,\dots,i_r)=D(i_1)\cup\dots\cup D(i_r).
\]
Then the $r$-th order bound is defined as
\[
	\delta_r(m)=\min\{\# D(i_1,\dots,i_r):i_1,\dots,i_r>m\}.
\]
This definition was introduced in \cite{HePe}.
It is proved by Farr\'an and Munuera in \cite{FaMu}
that for each numerical semigroup $\Lambda$ and each integer $r\geq 2$ there exists a constant $E_r=E(\Lambda,r)$, called $r$-th Feng-Rao number, such that
\begin{enumerate}
\item $\delta_r(m)=m+2-g+E_r$ for all $m$ such that $\lambda_m\geq 2c-2$ \cite[Theorem 3]{FaMu},
\item $\delta_r(m)\geq m+2-g+E_r$ for any $m$ such that $\lambda_m\geq c$ \cite[Theorem 8]{FaMu},
\end{enumerate}
where $c$ and $g$ are respectively the conductor and the genus of $\Lambda$.
Note that this is an extension of the Goppa bound for the case $r=1$, with $E_r=0$ \cite[Theorem 5.24]{HoLiPe}.

Furthermore, $E_r$ satisfies
\begin{enumerate}
\setcounter{enumi}{2}
\item $r\leq E_r\leq \lambda_{r-1}$ if $g>0$ (and $r\geq 2$) \cite[Proposition 5]{FaMu},
\item $E_r=\lambda_{r-1}$ if $r\geq c$ \cite[Proposition 5]{FaMu},
\item $E_r=r-1$ if $g=0$.
\end{enumerate}
Some further results related to the Feng-Rao number can be found 
in \cite{FaMu,FGL,DFGL}. Here we use the main result in the previous section to
obtain a lower bound on $E_r$, which is strictly better than the 
bound $E_r\geq r$ for $r > 2$ and for semigroups with more than two intervals of gaps.

\subsection{Bound on the Feng-Rao numbers}

For our bound on the Feng-Rao numbers we first need the next lemma.

\begin{lemma}
\label{l:maxj}
Consider the set of sets 
\begin{eqnarray*}
	{\mathcal A(a_1,a_r,r,\ell)}&=&\{A \subset{\mathbb N}_0: \#A=r,\\
&&\min(A)=a_1, \max(A)=a_r, \\
&&A \mbox{ contains at least }\ell
\mbox{ consecutive integers}\}.
\end{eqnarray*}
For each $A\in {\mathcal A}$ define ${\alpha}(A)=\max\{a\in A: a-\ell+1,\dots,a\in A\}$.
If $A$ has minimum ${\alpha}(A)$ among the sets in ${\mathcal A}$, then
\[
	{\alpha}(A)=\max\{a_1+\ell-1,a_1+(\ell-1)(a_1-a_r)+\ell(r-1)\}.
\]
\end{lemma}

\begin{IEEEproof}
Suppose that $A$ has minimum ${\alpha}(A)$ among the sets in ${\mathcal A}$.
If $a_1,a_1+1,\dots,a_1+(\ell-1)\in A$ and $\alpha(A)=a_\ell=a_1+\ell-1$, this means that there must be at least $\frac{r-\ell}{\ell-1}$ integers in 
the interval $[a_1,a_r]$ not belonging to $A$ since for each 
$\ell -1$ integers remaining in $A$ there must be at least one element not in $A$.
But the number of integers in $[a_1,a_r]\setminus A$ is 
$a_r-a_1+1-r$. So, $a_r-a_1+1-r\geq \frac{r-\ell}{\ell-1}$
or, equivalently, $\ell-1\geq (\ell-1)(a_1-a_r)+\ell(r-1).$
Hence, $\alpha(A)=a_1+\ell-1=\max\{a_1+\ell-1,a_1+(\ell-1)(a_1-a_r)+\ell(r-1)\}$.

Otherwise, we can assume that $\alpha(A)>a_1+(\ell-1)$. 
In this case, $A$ must be equal to
\[
	\begin{split}
	&\{a_1,a_1+1,a_1+2,\dots,\alpha(A)=a_r-\ell t\}\cup\{a_r-\ell t+2,\dots,a_r-\ell(t-1)\}\\
	&\quad\cup\dots\cup\{a_r-2\ell+2,\dots,a_r-\ell\}\cup\{a_r-\ell+2,\dots,a_r\},
	\end{split}
\]
for $t$ the number of integers in the interval $[a_1,a_r]$ not belonging to $A$, that is, $t=a_r-a_1+1-r$. So, $\alpha(A)=a_1+(\ell-1)(a_1-a_r)+\ell(r-1)=\max\{a_1+\ell-1,a_1+(\ell-1)(a_1-a_r)+\ell(r-1)\}.$
\end{IEEEproof}

\begin{theorem}\label{t:boundEr}
Suppose that $\ell>1$ is an integer and that $n_{\ell-1}$ is the number of intervals of at least $\ell-1$ gaps of $\Lambda$. Then the following inequality holds.
\begin{eqnarray}\label{boundlb}
E_r&\geq&\min\left\{r-2+\left\lceil\frac{r}{\ell-1}\right\rceil, r-1+\left\lceil\frac{(\ell-1)n_{\ell-1}}{\ell}\right\rceil\right\}.
\end{eqnarray}
\end{theorem}

\begin{IEEEproof}
By definition of $\delta_r(m)$, there exist integers $i_1,\dots,i_r$ with $m<i_1<\dots<i_r$
such that $\delta_r(m)=\#D(i_1,\dots,i_r)$. The integers $i_1,\dots,i_r$ minimize $\#D(i_1,\dots,i_r)$.
Denote $A$ the set $\{i_1,\dots,i_r\}$.
Suppose that $m$ is an integer with $m\geq 2c-1-g$.
By the definition of $E_r$, $\delta_r(m)=m+2-g+E_r$.

Since $A$ minimizes $\#D(i_1,\dots,i_r)$, it necessarily holds that $i_1=m+1$. Applying Theorem~\ref{t:maxidealgap} to the ideal $\Lambda\setminus D(i_1,\dots,i_r)$, we get $(m+2-g+E_r)+(2g-1)\geq \lambda_{i_r }=g+i_r$. 
Reorganizing the inequality gives 
\begin{equation}\label{boundEr}
	i_r\leq m+1+E_r.
\end{equation}
Suppose now that there are no $\ell$ consecutive integers in $A$. Then 
\begin{equation}
\label{eq:dos}
i_r\geq m+1+r-1+\left\lceil\frac{r-(\ell-1)}{\ell-1}\right\rceil.
\end{equation}
Now, by \eqref{boundEr}, $E_r\geq r-2+\left\lceil\frac{r}{\ell-1}\right\rceil$. Suppose on the other hand that there are at least $\ell$ consecutive integers in $A$. Let $i_j$ be the maximum 
integer in $A$ such that $i_j-\ell+1,\dots,i_j\in A$ and so 
$i_{j-\ell+1}=i_j-\ell+1,\dots,i_{j-1}=i_j-1$ and
\[
	\lambda_{i_{j-\ell+1}}=\lambda_{i_j}-\ell+1,\dots,\lambda_{i_{j-1}}=\lambda_{i_j}-1.
\]
Let
\[
	\Gamma=\{\lambda\in\Lambda: \lambda+1,\dots,\lambda+\ell-1\not\in\Lambda\}.
\]
In particular, if $\lambda\in \Gamma$ then $\lambda< c$, for $c$ the conductor of $\Lambda$.
Obviously $\#\Gamma=n_{\ell-1}$. If $\lambda\in \Gamma$ then  
\begin{eqnarray*}
(\lambda_{i_j}-1)-\lambda &\in& D(i_{j-1})\setminus D(i_j),\\ 
(\lambda_{i_j}-2)-\lambda &\in& D(i_{j-2})\setminus D(i_j), \\
&\vdots&\\
(\lambda_{i_j}-\ell+1)-\lambda &\in& D(i_{j-\ell+1})\setminus D(i_j).
\end{eqnarray*}
and so 
\[
	\{\lambda_{i_j}-1-\lambda, \lambda_{i_j}-2-\lambda, \dots, \lambda_{i_j}-\ell+1-\lambda\} \subseteq D(i_{j-\ell+1},\dots,i_{j-1})\setminus D(i_j).
\]

In fact, 
\[
	\cup_{\lambda\in \Gamma}\{\lambda_{i_j}-1-\lambda, \dots, \lambda_{i_j}-\ell+1-\lambda\} \subseteq D(i_{j-\ell+1},\dots,i_{j-1})\setminus D(i_j)
\]
and the sets in this union are disjoint.
Indeed, for $\lambda,\lambda'\in \Gamma$, with $\lambda>\lambda'$, it holds $\lambda-\lambda'\geq \ell$. Then, 
$\min\{\lambda_{i_j}-1-\lambda', \dots, \lambda_{i_j}-\ell+1-\lambda'\}=
\lambda_{i_j}-\ell+1-\lambda'\geq \lambda_{i_j}+1-\lambda>
\max\{\lambda_{i_j}-1-\lambda, \dots, \lambda_{i_j}-\ell+1-\lambda\}$.

So, 
\begin{eqnarray}
\#D(i_1,\dots,i_r)&\geq& \#D(i_{j-\ell+1},\dots,i_j)
\label{eq:risj}\\
&\geq &
(\ell-1)n_{\ell-1}+\nu_{i_j}\nonumber\\
&=&(\ell-1)n_{\ell-1}+i_j+1-g.\nonumber
\end{eqnarray}
Since $D(i_1,\dots,i_r)=m+2-g+E_r$ we get that $m+2-g+E_r\geq (\ell-1)n_{\ell-1}+i_j+1-g$,
so 
\begin{equation}
\label{eq:first}
E_r\geq (\ell-1)n_{\ell-1}+i_j-m-1.
\end{equation}
Now, 
by the maximality of $j$, and by Lemma~\ref{l:maxj},
\begin{equation}
\label{eq:tres}
i_j\geq \max\{i_1+\ell-1,i_1+(\ell-1)(i_1-i_r)+\ell(r-1)\}.
\end{equation}
This implies
\begin{equation}
\label{eq:maxa}
i_j\geq i_1+\ell-1,
\end{equation}
and
\begin{equation}
\label{eq:maxb}
i_j\geq i_1+(\ell-1)(i_1-i_r)+\ell(r-1).
\end{equation}

On one hand, using \eqref{eq:first} and 
\eqref{eq:maxa}, 
we deduce that $E_r\geq(\ell-1)(n_{\ell-1}+1)$.
On the other hand, using \eqref{eq:first} and \eqref{eq:maxb},  and then \eqref{boundEr},
\begin{eqnarray*}
E_r&\geq& (\ell-1)n_{\ell-1}+i_1+(\ell-1)(i_1-i_r)+\ell(r-1)-m-1
\\&=&(\ell-1)n_{\ell-1}+(\ell-1)(i_1-i_r)+\ell(r-1)
\\&\geq&
(\ell-1)n_{\ell-1}-(\ell-1)E_r+\ell(r-1)
\end{eqnarray*}
and we conclude
that $E_r\geq r-1+\left\lceil\frac{(\ell-1)n_{\ell-1}}{\ell}\right\rceil$.

We have seen that, depending on whether $I$ contains $\ell$ consecutive integers or not,
either $E_r\geq r-2+\left\lceil\frac{r}{\ell-1}\right\rceil$ or 
$E_r\geq \max\{(\ell-1)(n_{\ell-1}+1),
r-1+\left\lceil\frac{(\ell-1)n_{\ell-1}}{\ell}\right\rceil\}$.
So, 
we deduce the bounds
\begin{eqnarray*}
E_r&\geq&\min\{r-2+\left\lceil\frac{r}{\ell-1}\right\rceil, (\ell-1)(n_{\ell-1}+1)\},
\\
E_r&\geq&\min\{r-2+\left\lceil\frac{r}{\ell-1}\right\rceil, r-1+\left\lceil\frac{(\ell-1)n_{\ell-1}}{\ell}\right\rceil\}.
\end{eqnarray*}

Notice, though, that the second bound is always at least as good as the first one, so the first one can be ommitted. Indeed,
if $
r-2+\left\lceil\frac{r}{\ell-1}\right\rceil
\leq
r-1+\left\lceil\frac{(\ell-1)n_{\ell-1}}{\ell}\right\rceil 
$, then we are done. On the contrary, assume that 
$
r-2+\left\lceil\frac{r}{\ell-1}\right\rceil
>
r-1+\left\lceil\frac{(\ell-1)n_{\ell-1}}{\ell}\right\rceil $. We need to prove that in this case 
$r-1+\left\lceil\frac{(\ell-1)n_{\ell-1}}{\ell}\right\rceil\geq
(\ell-1)(n_{\ell-1}+1)$.

If $
r-2+\left\lceil\frac{r}{\ell-1}\right\rceil
>
r-1+\left\lceil\frac{(\ell-1)n_{\ell-1}}{\ell}\right\rceil 
$
then 
$\left\lceil\frac{r}{\ell-1}\right\rceil
>
1+\left\lceil\frac{(\ell-1)n_{\ell-1}}{\ell}\right\rceil $
which implies that
$
\frac{r}{\ell-1}
>
1+\frac{(\ell-1)n_{\ell-1}}{\ell}
$ 
and so 
 $r>(\ell-1)(1+\frac{(\ell-1)n_{\ell-1}}{\ell})=
(\ell-1)((n_{\ell-1}+1)-\frac{n_{\ell-1}}{\ell}).$
This implies 
$r+\frac{(\ell-1)n_{\ell-1}}{\ell}>(\ell-1)(n_{\ell-1}+1)$ and so
$r-1+\left\lceil\frac{(\ell-1)n_{\ell-1}}{\ell}\right\rceil\geq
(\ell-1)(n_{\ell-1}+1)$
as desired.
\end{IEEEproof}

\begin{remark}
Notice that if $r\leq 2(\ell-1)$ then the bound in Theorem~\ref{t:boundEr}
does not improve the bound $E_r\geq r$.
So, the bound makes sense when $\ell<r/2+1$.
The same happens for $n_{\ell-1}=0$. So, we are interested in the values of $\ell$ such that
\begin{itemize}
\item $n_{\ell-1}>0$
\item $\ell<r/2+1$.
\end{itemize}
\end{remark}

\begin{corollary}
\label{c:bound}
Let $m$ be such that $\lambda_m\geq c$ and let $\ell\geq 2$. Then
\[
	\delta_r(m)\geq m+2-g+
\min\{r-2+\left\lceil\frac{r}{\ell-1}\right\rceil, r-1+\left\lceil\frac{(\ell-1)n_{\ell-1}}{\ell}\right\rceil\}.
\]
\end{corollary}
 
\begin{remark}
From bound \eqref{boundlb}, 
taking $\ell=2$, we deduce that, if $n$ is the number of intervals of (at least one) gaps of $\Lambda$, then
\begin{equation}
	E_r\geq \min\{2(r-1),r-1+\lceil n/2\rceil\}.
\label{boundu}
\end{equation}
\end{remark}

\begin{remark}
  If $r=2$ or $n\leq 2$ then bound~\eqref{boundu} equals 
the bound $E_r\geq r$.
But in any other case, bound~\eqref{boundu} is better.
\end{remark}

\begin{corollary}
If $\Lambda$ is a semigroup with conductor $c$ and $n$ intervals of gaps then, for any $m$ with $\lambda_m\geq c$,
$$\delta_r(m)\geq \left\{\begin{array}{ll}m-g+2r&\mbox{ if }r\leq \lceil n/2\rceil +1,\\
m-g+r+\lceil n/2\rceil+1&\mbox{ otherwise.}
\end{array}\right.$$
\end{corollary}

\subsection{Sharpness of the bound}

Analyzing the proof of Theorem~\ref{t:boundEr} we see that
the bound~\eqref{boundlb} may be sharp only if 
\begin{enumerate}
\item
The inequality in \eqref{boundEr}, obtained applying Theorem \ref{t:maxidealgap}
to the ideal $\Lambda\setminus D(i_1,\dots i_r)$,
is indeed an equality. This means, 
by applying Theorem~\ref{t:attain} 
to the same ideal, that
$D(i_1,\dots,i_r)=D(i_r)$, and so $i_1,\dots,i_{r-1}\subseteq i_r-\Lambda$. In particular, $i_r-i_{r-1}\geq \lambda_1$.
\item
Either the inequality in \eqref{eq:dos} or both the inequalities in \eqref{eq:risj} and \eqref{eq:tres} are indeed equalities,
which means that the difference between $i_r$ and $i_{r-1}$ is at most two. So, $i_r-i_{r-1}\leq 2$.
\end{enumerate}

We conclude that the only semigroups for which the bound may be sharp are 
hyperelliptic semigroups, that is, semigroups that contain $2$.

It is proved in \cite[Theorem 1]{FGL} that for hyperelliptic semigroups,
$E_r=\lambda_{r-1}=2(r-1).$
The bound~\eqref{boundlb} 
for the hyperelliptic semigroup of genus $g$ is
$$E_r\geq\left\{\begin{array}{ll}
r-1 &\mbox{if }\ell>2\\
2(r-1)&\mbox{if }\ell=2\mbox{ and }r-1\leq\lceil g/2\rceil\\
r-1+\lceil g/2\rceil&\mbox{if }\ell=2\mbox{ and }r-1>\lceil g/2\rceil\\
\end{array}\right.$$
Hence the bound is sharp if and only if $\Lambda$ is hyperelliptic, $\ell=2$, and 
$r\leq1+\lceil g/2\rceil$.

\subsection{An example}
As an example consider the semigroup $$\{0,3,6,9,\dots,36,37,38,\dots\},$$
with $\ell= 3$.
Let us analyze the bounds in \eqref{boundlb} and
\eqref{boundu} for different values of $r$.
In this case $n_{\ell-1}=n_1=12$ and so
the bound in \eqref{boundlb} is
$$\min\{r-2+\left\lceil\frac{r}{2}\right\rceil, r+7\}$$ while
the bound in \eqref{boundu} is
$$\min\{2(r-1),r+5\}.$$

\subsubsection*{Case $r=6$}
Bound \eqref{boundlb} is $\min\{7,13\}=7$
while bound \eqref{boundu} is
$\min\{10,11\}=10$.
So, bound \eqref{boundu} (with the first element being the minimum) is better than bound \eqref{boundlb}.

\subsubsection*{Case $r=8$}
Bound \eqref{boundlb} is $\min\{10,15\}=10$
while bound \eqref{boundu} is $\min\{14,13\}=13$
So, bound \eqref{boundu} (with the second element being the minimum) is better than bound \eqref{boundlb}.

\subsubsection*{Case $r=15$}
Bound \eqref{boundlb} is $\min\{21,22\}=21$
while bound \eqref{boundu} is $\min\{28,20\}=20$
So, bound \eqref{boundlb} (with the first element being the minimum) is better than bound \eqref{boundu}.

\subsubsection*{Case $r=20$}
Bound \eqref{boundlb} is $\min\{28,27\}=27$
while bound \eqref{boundu} is $\min\{38,25\}=25$
So, bound \eqref{boundlb} (with the second element being the minimum) is better than bound \eqref{boundu}.

\section{Examples of the computation of the number of intervals of gaps}

Now we analyze $n_\ell$ for two classical families of codes, that is, for Hermitian codes and for codes in one of the Garcia-Stichtenoth's towers of codes attaining the Drinfeld-Vl\u adu\c t bound, as well as their respective generalizations to semigroups generated by intervals and inductive semigroups.

\subsection{Hermitian codes}
Let $q$ be a prime power.  The Hermitian curve over ${\mathbb F}_{q^2}$ is defined by the affine equation $$x^{q+1}=y^{q}+y$$ and it has a single rational point at infinity and $q^3$ more rational points.  Its weight hierarchy has already been studied in \cite{YaKuSt,BarMun}.  However, for its simplicity, we wanted to give a description of $n_\ell$.  The Weierstrass semigroup at the rational point at infinity is generated by $q$ and $q+1$ \cite{Stichtenoth:hermite,HoLiPe}.  Some results concerning this semigroup can be found in \cite{BrOS:hermite} and, in particular, concerning the weight hierarchy, in \cite{DFGLtwogen}.

The semigroup generated by $q$ and $q+1$ is $\{0\}\cup\{q,q+1\}\cup\{2q,2q+1,2q+2\}\cup\dots\cup\{(q-2)q,(q-2)q+1,\dots,(q-2)q+(q-2)=(q-1)q-2\}\cup\{j\in{\mathbb N}_0: j\geq (q-1)q\}$.
It is easy then to see that the lengths of the intervals of gaps, as they appear in the semigroup, are $q-1,q-2,\dots,1$. So, 
$$
n_\ell=\left\{
\begin{array}{ll}
q-\ell &\mbox{ if }1\leq \ell\leq q\\
0&\mbox{ if }\ell\geq q
\end{array}\right.
$$

\subsection{A generalization: semigroups generated by intervals}

The semigroup of the Hermitian curve can be thought as generated by the interval of length $2$ starting at $q$.
Suppose that a numerical semigroup is generated by the interval of $x$ integers starting at $a$: $\{a,a+1,\dots,a+x-1\}$. 
These semigroups
can be found, for instance, in \cite{intervals}. Also, the Feng-Rao numbers of such semigroups are studied in \cite{DFGL}.

In this case, the semigroup is 
$\{0\}\cup\{a,a+1,\dots,a+x-1\}\cup\{2a,2a+1,\dots,2a+2x-2\}\cup\dots\cup\{ka,\dots,ka+kx-k\}\cup\{(k+1)a,\dots,(k+1)a+(k+1)x-(k+1)\}\cup\dots$.

The gap intervals correspond to the sets between $ka+kx-k+1$ and $(k+1)a-1$ for $k\geq 0$ and whenever $(k+1)a-1\geq ka+kx-k+1$.  The number of gaps of these sets is $(a-1)-k(x-1)$.  So, 
\begin{eqnarray*}
n_\ell&=&\#\left\{k \mbox{ such that } \left\{\begin{array}{l}(a-1)-k(x-1)\geq \ell\\k\geq 0\end{array} \right.\right\}\\
&=& \#\left\{k \mbox{ such that } 0\leq k\leq \frac{a-1-\ell}{x-1}\right\}\\
&=& \left\{\begin{array}{ll}\left\lfloor\frac{a-1-\ell}{x-1}\right\rfloor+1
&\mbox{ if }1\leq \ell\leq a\\
0
&\mbox{ if }\ell\geq a\\
\end{array}\right.
\end{eqnarray*}

We see that this result generalizes the one previously found for Hermitian codes.
We leave it as an open problem to compare the bound proved in Theorem~\ref{t:boundEr}, using this value of $n_\ell$ with the results in \cite{DFGL}.

\subsection{Codes on the Garcia-Stichtenoth tower of codes}

Garcia and Stichtenoth gave in \cite{GaSt:tff} a celebrated tower of function fields attaining the Drinfeld-Vl\u adu\c t bound, which became of great importance in the area of algebraic coding theory.  Since then other towers have also been found, although we will focus on the tower in \cite{GaSt:tff}. It is defined over the finite field with $q^2$ elements ${\mathbb F}_{q^2}$ for $q$ a prime power. It is given by ${\mathcal F}_{1}={\mathbb F}_{q^2}(x_1)$; ${\mathcal F}_{m}={\mathcal F}_{m-1}(x_m)$, with $x_m$ satisfying $$x_m^q+x_m=\frac{x_{m-1}^{q}}{x_{m-1}^{q-1}+1}.$$ It is shown in \cite{GaSt:tff} that the number of its rational points is $N_q({\mathcal F}_{m})\geq(q^2-q)q^{m-1}$ and that the genus $g_m$ of ${\mathcal F}_{m}$ is $g_m= (q^{\lfloor\frac{m+1}{2}\rfloor}-1) (q^{\lceil\frac{m-1}{2}\rceil}-1).  $ Hence, the ratio between the genus $g({\mathcal F}_{m})$ and $N_{q^2}({\mathcal F}_{m})$ converges to $1/(q-1)$, the Drinfeld-Vl\u adu\c t bound, as $m$ increases.  From these curves one can construct asymptotically good sequences of codes.

For every function field ${\mathcal F}_{m}$ in the tower we distinguish the rational point $Q_{m}$ that is the unique pole of $x_1$.  The Weierstrass semigroup $\Lambda_{m}$ at $Q_{m}$ in ${\mathcal F}_{m}$ was recursively described in \cite{PeStTo}.  \index{Garcia-Stichtenoth towers!semigroup} Indeed, the semigroups are given recursively by
 \begin{equation}
 \label{eq:semigroups}
 \begin{array}{rcl}
 \Lambda_{1}&=&{\mathbb N}_0
 \\
 \Lambda_{m}&=&q\cdot\Lambda_{m-1}\cup\{i\in{\mathbb N}_0: i\geq
 q^m-q^{\lfloor\frac{m+1}{2}\rfloor}\}.
 \end{array}
 \end{equation}
In \cite{BrOS:towers} 
a non-recursive description of these semigroups is given as follows.
\begin{equation}
\label{eq:nonrecursive}
\Lambda_{m}=\bigsqcup_{i=1}^{\lfloor\frac{m}{2}\rfloor}q^{m-2i+1}A_{i}\sqcup\{j\in{\mathbb N}_0:
j\geq c_m\},
\end{equation}
where
$c_m$ is the conductor of $\Lambda_{m}$, which is $q^m-q^{\lfloor\frac{m+1}{2}\rfloor}$, and
 $A_i=\{c_{2i-1}+j: j=0,\dots,q^{i-1}(q-1)-1\}$.

From \eqref{eq:nonrecursive} we can deduce that there are exactly $\# A_i=q^{i-1}(q-1)$ intervals of length $q^{m-2i+1}-1$.  Now, if $j$ is maximum such that  $\ell \leq q^{m-2j+1}-1$ then 
$n_\ell=\sum_{i=1}^{j}q^{i-1}(q-1)=q^{j}-1$. But $\ell \leq q^{m-2j+1}-1$ is equivalent to 
$j\leq \frac{m+1-\log_q(\ell+1)}{2}$ and we can take
$j=\lfloor\frac{m+1-\log_q(\ell+1)}{2}\rfloor$.
So, 
\begin{equation}
\label{eq:tor}
n_\ell=q^{\lfloor\frac{m+1-\log_q(\ell+1)}{2}\rfloor}-1.
\end{equation}

\subsection{Inductive semigroups}

In \cite{PeTo} an inductive sequence of semigroups is defined as a sequence for which there exist sequences $(a_m: m \in {\mathbb N})$ and $(b_m: m \in {\mathbb N})$, with $a_mb_{m}\leq b_{m+1}$ such that $\Lambda_1={\mathbb N}_0$
and $\Lambda_m=a_m\Lambda_{m-1}\cup \{n\in {\mathbb N}_0 : n\geq a_mb_{m}\}$ or all $m>1$.
See also \cite{CaFaMu,sparse}.

The semigroups in the previous subsection are an example of inductive sequence of semigroups with $a_m=q$ for all $m$.
In general, if $a_m=q$ for all $m$, then the semigroup $\Lambda_m$ equals
the disjoint union of the next sets (recall the condition 
$qb_{m}\leq b_{m+1}$ for all $m>1$).
\begin{eqnarray*}
\Lambda_m^{(q^{m-1})}&=&\{q^{m-1}, 2q^{m-1}, \dots, b_2 q^{m-1}\},\\
\Lambda_m^{(q^{m-2})}&=&\{b_2q^{m-1}+q^{m-2}, b_2q^{m-1}+2q^{m-2}, \dots, b_3 q^{m-2}\},\\
\Lambda_m^{(q^{m-3})}&=&\{b_3q^{m-2}+q^{m-3}, b_3q^{m-2}+2q^{m-3}, \dots, b_4 q^{m-3}\},\\
&\vdots&\\
\Lambda_m^{(q)}&=&\{b_{m-1}q^{2}+q, b_{m-1}q^{2}+2q, \dots, b_m q\},\\
\Lambda_m^{(1)}&=&\{b_m q +1, b_mq+2,\dots\}.
\end{eqnarray*}
So, $\Lambda_m$ has $b_2$ intervals of $q^{m-1}-1$ gaps, 
$\frac{b_3q^{m-2}-b_2q^{m-1}}{q^{m-2}}=b_3-qb_2$ intervals of $q^{m-2}-1$ gaps, 
$\frac{b_4q^{m-3}-b_3q^{m-2}}{q^{m-3}}=b_4-qb_3$ intervals of $q^{m-3}-1$ gaps, 
and so on.
In general, it has 
$b_{k+1}-qb_k$ intervals of $q^{m-k}-1$ gaps, for $k\geq 1$, where $b_1$ may be assumed to be $0$.

Now, when looking for intervals with at least $\ell$ consecutive gaps,
we need to take into account that $q^{m-k}-1\geq \ell$ if and only if 
$k\leq m-\log_q(\ell +1)$. Let $N=\lfloor m-\log_q(\ell +1)\rfloor$. Then, 
\begin{eqnarray}n_\ell&=&\sum_{k=1}^{N}b_{k+1}-qb_{k}\nonumber
\\&=&b_{N+1}+\sum_{k=1}^{N}(1-q)b_k.\label{eq:ind}
\end{eqnarray}
Let us check that this result generalizes \eqref{eq:tor}.
In fact, for the inductive semigroups in the previous section, one has
$b_m=q^{m-1}-q^{\lfloor\frac{m-1}{2}\rfloor}$. Substituting this value in \eqref{eq:ind}
we get
\begin{eqnarray*}
n_\ell&=& q^N-q^{\lfloor\frac{N}{2}\rfloor}+(1-q)\left(\sum_{k=1}^N(q^{k-1}-q^{\lfloor\frac{k-1}{2}\rfloor})\right)\\
&=& q^N-q^{\lfloor\frac{N}{2}\rfloor}-(q-1)\left(\sum_{k=1}^Nq^{k-1}\right)+(q-1)\left(\sum_{k=1}^N
q^{\lfloor\frac{k-1}{2}\rfloor}\right)\\
&=& q^N-q^{\lfloor\frac{N}{2}\rfloor}-(q-1)\frac{q^N-1}{q-1}
+(q-1)\left(\sum_{k=1}^Nq^{\lfloor\frac{k-1}{2}\rfloor}\right)\\
&=& 1-q^{\lfloor\frac{N}{2}\rfloor}
+(q-1)\left(\sum_{k=1}^Nq^{\lfloor\frac{k-1}{2}\rfloor}\right).\\
\end{eqnarray*}

If $N$ is even then 
\begin{eqnarray*}(q-1)\left(\sum_{k=1}^Nq^{\lfloor\frac{k-1}{2}\rfloor}\right)&=&
2(q-1)(1+q+q^2+\dots+q^{\frac{N}{2}-1})\\&=&
2(q-1)\frac{q^{N/2}-1}{q-1}\\&=&
2(q^{N/2}-1),
\end{eqnarray*}
while, 
if $N$ is odd, then 
\begin{eqnarray*}(q-1)\left(\sum_{k=1}^Nq^{\lfloor\frac{k-1}{2}\rfloor}\right)&=&
2(q-1)(1+q+q^2+\dots+q^{\frac{N-1}{2}-1}) + (q-1) q^{\frac{N-1}{2}}\\&=&
2(q-1)\frac{q^{\frac{N-1}{2}-1}}{q-1} + (q-1) q^{\frac{N-1}{2}}\\&=&
2(q^{\frac{N-1}{2}}-1) + q^{\frac{N+1}{2}}-q^{\frac{N-1}{2}}\\&=&
q^{\frac{N+1}{2}}+q^{\frac{N-1}{2}}-2.\\
\end{eqnarray*}
In both cases, we obtain that $n_\ell=q^{\lfloor\frac{N+1}{2}\rfloor}-1$. Now, substituting $N$ by its value, we check that 
$n_\ell=q^{\lfloor\frac{\lfloor m+1-\log_q(\ell +1)\rfloor}{2}\rfloor}-1$.
The floor in the numerator of the exponent is redundant, and so this result coincides with \eqref{eq:tor}.


\begin{IEEEbiographynophoto}{Maria Bras-Amorós} received her PhD in Applied Mathematics in 2003 from Universitat Politècnica de Catalunya and part of her doctoral and postdoctoral work was developed at San Diego State University, California. She is a tenure-track associate professor at Universitat Rovira i Virgili in Tarragona. She has been with Universitat Politècnica de Catalunya and Universitat Autònoma de Barcelona. Her main research interests are in the area of coding theory and discrete mathematics. \end{IEEEbiographynophoto}

\begin{IEEEbiographynophoto}{Kwankyu Lee} received the B.Sc., M.Sc., and Ph.D.~degrees in Mathematics in 1998, 2000, and 2005 respectively, from Sogang University in Korea. He is an associate professor in the Department of Mathematics and Education at Chosun University in Gwangju, Korea. His research interests include algebraic coding theory, cryptography, and discrete mathematics.
\end{IEEEbiographynophoto}

\begin{IEEEbiographynophoto}{Albert Vico-Oton} received his PhD in Computer Science in 2013 from the University Rovira i Virgili, Catalonia, Spain. He is currently a member of the engineering team at Midokura, working on the innovative Midokura's software defining networking solutions for private clouds and data center management. He also worked as the team leader of the Distributed Applications and Networks Area - DANA http://dana.i2cat.net at the i2CAT Foundation research center of Catalonia, where he leaded different network research projects based on Future Internet with special focus on SDN technologies. His research interests include software defined networking, network function virtualization, coding theory, and discrete mathematics.\end{IEEEbiographynophoto}

\end{document}